# HAT-P-32 b: what can be deduced from transit observations in Hα and He I lines?

Shaikhislamov I.F.[1,2,3], Sharipov S.S.[1], Khodachenko M.L.[4], Miroshnichenko I.B.[1,3], Rumenskikh M.S.[1,2], Golubovsky M.P.[1,2], Berezutsky A.G.[1,2]

1) Institute of Laser Physics SB RAS, Novosibirsk, Russia
2) Institute of Astronomy, Russian Academy of Sciences, Moscow, Russia
3) Novosibirsk State Technical University, Novosibirsk, Russia
4) Institute for Space Research, Graz, Austria

Email address: ShaikhislamovIldar@yandex.ru

**ABSTRACT**

HAT-P-32 b is the first exoplanet for which absorption in hydrogen and helium lines has been measured simultaneously. In addition to the relatively large maximum depth of ~5% in both lines, observations have revealed very long pre-transit signatures. In this paper, we apply a 3D aeronomy model to simulate the detailed spectrally resolved absorption profiles at mid-transit, formed in the upper atmosphere, as well as the integral transit curves formed by atmospheric material accumulating around the star. By fitting the observations, we derive key atmospheric characteristics, including the atmospheric composition and metallicity, and constrain the stellar XUV, hard X-ray, and Lyα fluxes, as well as the intensity of the stellar plasma wind. We show that the presence of metals crucially affects the absorption in hydrogen and helium lines by accelerating the dissociation of $H_2$. We also demonstrate that several processes not previously included in the modeling of HAT-P-32 b are important for the interpretation of the observations, such as the 3D asymmetric structure of the atmosphere, heating and cooling by excited states of hydrogen and metals, and the stellar VUV flux.

## Introduction

HAT-P-32 is a unique system because it provides, for the first time, a robust measurement of simultaneous absorption in hydrogen and helium in a planetary atmosphere (Czesla et al. 2022). The depth of the excess absorption at mid-transit exceeds 5% in the Hα (656.3 nm) and He I (1083 nm) lines, with an equivalent width of ~100 mÅ. Such values imply that HAT-P-32 b is a hot Jupiter with a highly inflated atmosphere losing mass at a rate of ~$10^{13}$ g s$^{-1}$. The system therefore poses a challenge for hydrodynamic models aiming to derive reliable quantitative conclusions from the observations. It represents an excellent test case for validating models that attempt to reproduce transit absorption simultaneously in different spectral lines of different elements.

Czesla et al. 2022 provide a detailed summary of the HAT-P-32 system based on previous studies, which we also use in the present work. Observations with the CARMENES spectrograph at the Calar Alto Observatory were obtained on 1 and 9 December 2018. On 30 August 2019, XMM-Newton observed HAT-P-32, revealing a high X-ray flux. The RV timing of the absorption signals in the Hα and He I lines clearly indicates their origin in the planetary atmosphere. In addition to significant absorption at the line cores at mid-transit, the red wings of both lines

([+15, +50] km s$^{-1}$ in Doppler velocity) show a prominent early ingress, starting about one hour before the first optical contact. At the same time, no apparent egress after the transit was detected.

Next observation of HAT-P-32 was made by Zhang et al. 2023 on the Hobby-Eberly Telescope during 11 nights at August-November of 2020 yielding the helium excess absorption exceeding 8%. The light curve in the He I 1083 nm line (hereafter referred to as simply the He I line), integrated in the stellar rest frame over the velocity range ±40 km s$^{-1}$, indicates that the escaping atmosphere spans a projected length exceeding 53 planetary radii. The extended leading tail produces detectable absorption up to 4 hours before the optical transit, while the trailing tail is significantly shorter. These observations suggest that, in addition to the inflated atmosphere immediately surrounding the planet—responsible for the maximum absorption of about 5–8% in the Hα and He I lines at mid-transit—there is likely a time-variable component of more rarefied but still sufficiently dense atmospheric material extending along the orbit with a complex asymmetric geometry.

To explain the observations, Czesla et al. 2022 used two different 1D hydrodynamic models, applied separately to the He I and Hα lines. The hydrogen model (García Muñoz & Schneider 2019) solves the NLTE kinetic equations using XUV and VUV stellar fluxes. Helium is not included in this hydrogen model. For the parameters providing the best fit to the Hα absorption profile, the model produces a narrow spatial ring of H I($n$=2) population with a width of 0.1R$_p$ located at about 1.8R$_p$. At the first Lagrange point, the temperature and velocity reach values of 1.5×10$^4$ K and 15 km s$^{-1}$, respectively. The calculated mass-loss rate is 1.3×10$^{13}$ g s$^{-1}$. For the He I line, a version of the isothermal Parker wind model combined with calculation of metastable level kinetics (Lampón et al. 2020) was used, with temperature and mass-loss rate treated as free parameters. A fit with parameters corresponding to the Hα modeling is obtained at significantly sub-solar helium abundance of H/He = 99/1. The widths of the absorption profiles of both lines, comparable to the observations, are produced by plasma bulk motion and empirical microturbulence. Lampón et al. 2023 used the same model to fit the observations. The derived parameters are the mass-loss rate of 1.3×10$^{13}$ g s$^{-1}$, a temperature of 1.24×10$^4$ K, and an H/He ratio of 99/1. The solution shows a sharp ionization front at about 1.6R$_p$. It is rather degenerate with respect to combination of these parameters.

In Yan et al. 2024 a 1D NLTE code was employed to calculate the temperature, velocity, and densities in the atmosphere based on hydrogen–helium plasma photochemistry. The H I($n$=2) population was calculated based on the absorption of stellar Lyα photons, whose transport is treated using a Monte Carlo code. For the He I line, it was found that either helium abundance must be very small, H/He ≥ 99.5/0.5, at the XUV flux measured by XMM-Newton, or XUV flux must be ten times lower for the solar abundance H/He = 92/8. A simultaneous fit to both lines was achieved by (Yan et al. 2024) for H/He ≥ 99.5/0.5, with the mass-loss rate of about (1.0–3.1)×10$^{13}$ g s$^{-1}$ and a temperature of about 13000 K. Besides, the stellar Lyα flux must also be very high, approximately comparable to the total XUV flux and about two orders of magnitude higher than the solar value.

In Yan et al. 2024, the possible effect of metallicity was estimated for the first time for HAT-P-32 b. The role of metallicity was considered in terms of its effect on the mean molecular weight of the atmosphere. It was found that solar metallicity provides the best fit to the observations.

To analyze the pre-transit absorption features, 1D models are not suitable. Czesla et al. 2022 devised a phenomenological model consisting of a spherical absorbing shell and superimposed up-orbit stream directed towards the star. To fit the pre-transit features, a velocity of ~100 km s$^{-1}$ and a mass flow rate ~$10^6$ g s$^{-1}$ were required for the stream.

Zhang et al. 2023 used 3D hydrodynamic model to simulate absorbing tails in the He I line. The model is nearly isothermal but assumes different temperatures for the planet and the star. The equations are solved in a rotating spherical reference frame centered on the host star. The aeronomy of the atmosphere is not calculated explicitly, and the planetary wind is generated by adjusting the escape parameter. To calculate the ionization, recombination, and level kinetics of helium, the XUV spectrum of τ Boo was used, combined with modeled spectra at λ > 120 nm. The best fit to the mid-transit absorption depth was obtained for a mass-loss rate of $10^{12}$ g s$^{-1}$. It was shown that the planetary outflow forms extended tidal tails, both up-stream and down-stream, circumventing the star. In these tails, the absorbing material lies approximately in the stellar rest frame rather than in the planet's moving frame, consistent with the observational data. However, neither the depth of the pre-transit absorption nor its duration can be quantitatively reproduced by these simulations.

Interestingly, a similar system, HAT-P-67 b, exhibits planetary outflows extending over hundreds of planetary radii (Gully-Santiago et al. 2024) with exceptionally pronounced pre-transit feature. Modeling by 3D HD simulations of Nail et al. 2025 of both HAT-P-32 b and HAT-P-67 b have shown that the pre-transit absorption is generated by a leading outflow stream that propagates ahead of the planet and spirals toward the star.

Thus, summarizing previous modeling, the works that simulate both spectral lines rely on 1D models and involve a number of other limitations. Different codes are used to calculate the populations of excited hydrogen, helium, and the atmosphere structure in general. No cooling by metals was included, though it can be rather important at considered temperatures (Huang et al. 2017). The simulations start at relatively rarified layers of the atmosphere, corresponding to pressures of about 1 $\mu$bar. As a result, a very sharp temperature jump occurs at the planet boundary, from the equilibrium value to much higher temperatures due to intense XUV heating. The 3D simulations involving the stellar system do not consider the aeronomy of the atmosphere, generate the planetary wind empirically, and do not calculate absorption in the Hα line.

In our 3D hydrodynamic model, we do not make any fundamental simplifications and include as many aspects, as feasible. We investigate the role of metallicity, by taking into account photoionization of elements such as C, Mg, and Fe, which supply much more electrons than H and He at pressures below $10^{-3}$ bar. We also directly calculate the radiative cooling provided by these metals. We solve the kinetic equations for the excited hydrogen levels and for the metastable helium triplet He I($2^3$S). Radiative transfer calculations account for the attenuation of stellar XUV and VUV fluxes in the atmosphere. To take into account the scattering of Lyα photons in optically thick atmosphere, we employ an independent Monte Carlo code. To reproduce the pre- and post- transit features, we solve the 3D hydrodynamic equations in a global frame that includes the star and the stellar wind.

Above mentioned previous works on HAT-P-32 b have posed important questions that have to be analyzed further. Do the observations imply that the high hard X-ray (HXR) flux measured by XMM-Newton during a single session is required as well to explain the absorption depth measured at a very different date, taking into account its probable variability? Must the stellar Lyα flux be as high as 100 times the solar value to reproduce the observed absorption in the Hα line? Under what conditions in the upper atmosphere does the accumulation of material around the star take place, producing significant pre-transit absorption? Namely these questions are the motivation behind the present study.

In our previous paper on the HAT-P-32 b (Sharipov et al. 2025) we attempted to answer some of these questions, by combining 3D hydrodynamic and 1D Monte Carlo codes. It was shown that in a model describing the hydrogen atom with only a single excited level $n = 2$, the fit of absorption depth of Hα line at mid-transit can be achieved only for extreme Lyα fluxes of ~ 600 erg cm$^{-2}$ s$^{-1}$. Extending the model to include $n = 2$, 3, and 4 levels significantly changed the population of $n = 2$ state and the overall heating of the atmosphere, thereby relaxing the requirement on the Lyα flux to a physically feasible value of ~ 100 erg cm$^{-2}$ s$^{-1}$. It was also shown that metallicity is rather important for the structure of atmosphere, due to supplying photoelectrons in the dense layers. The best-fit solution corresponded to a metallicity of 10% of solar value, which already significantly affects the absorption in both lines. In this paper we make significant physically important steps further. We include cooling by metals by calculating several of the most important excited levels of C, O, and Mg atoms and first ionization stages. We show that a base atmosphere consisting of atomic hydrogen and helium above pressure of 0.1 bar produces extremely expanded atmosphere with spectral line shapes too strongly Doppler-broadened by outflowing streams to fit observations. On the other hand, a purely molecular $H_2$ atmosphere with helium is much more compact and produces too small absorption in the Hα line. The presence of metals at approximately solar abundance appears to be crucial because it facilitates rapid dissociation of $H_2$, resulting in a much better agreement between the simulated and observed absorption in both lines. Another important addition to the previous paper of Sharipov et al. 2025 is the simulation of atmosphere in a global reference frame including the star and the stellar wind in order to interpret the observations of Zhang et al. 2023 with significant pre-transit absorption in He I line.

The paper is organized as follows. Section 2 describes the model. Section 3 presents the results and is divided into three subsections: 3.1 (Atomic atmosphere), 3.2 (Molecular atmosphere), and 3.3 (Metals), each containing a table summarizing the corresponding simulation runs. Section 4 presents the results of modeling the atmospheric outflow in a global reference frame that includes the stellar system, together with calculations of the full transit absorption curves. The paper concludes with a summary of the main results.

## 2. Model description

The global 3D multi-fluid hydrodynamic (HD) model (hereafter referred to as Exo3D) used in the present work, has been already described in our earlier papers, e.g., in (Khodachenko et al. 2019, 2021a; Shaikhislamov et al. 2018, 2019, 2020a, 2020b). The code solves the hydrodynamic equations of continuity, momentum, and energy independently for each of the considered species in the multicomponent plasma. Among the considered species are H, He, as well as the heavier particles such as O, C, Mg, and Fe in atomic form and in ionization stages up to $Z = +2$. The model also includes molecular hydrogen species, such as $H_2$, $H_2^+$, and $H_3^+$ (Khodachenko et al. 2019; Shaikhislamov et al. 2018), together with the associated hydrogen chemistry.

Within the Exo3D code, we explicitly model the populations of excited hydrogen atoms by accounting for all relevant excitation and de-excitation processes. In the present work, we restrict the calculations to the levels $n = 2s$, $2p$, 3, and 4, following the approach and using the reaction rates reported by (García Muñoz & Schneider 2019). To properly account for the scattering of Lyα photons in the optically thick atmosphere, we complement the Exo3D simulations with Monte Carlo radiative transfer calculations. In the Exo3D code we approximately take it into account by introducing an empirical time of loss of Lyα photons as a function of the local trapping parameter (details are given in section 2.1). This approach was first tested by us for the ultra-hot Jupiter KELT-9 b (Shaikhislamov et al. 2025).

To derive the absorption in the He I line, we use the same code to calculate the population of metastable helium He I($2^3S$) as a separate fluid. The kinetic processes of He I($2^3S$) and the rates of the relevant reactions have been described in several of our previous papers, where the detected He I absorption in several exoplanets has been successfully reproduced, including WASP-107 b (Khodachenko et al. 2021a), GJ 3470 b (Shaikhislamov et al. 2020b), HD 209458 b (Khodachenko et al. 2021b), HD 189733 b (Rumenskikh et al. 2022), and TOI-421 b (Berezutsky et al. 2022).

The populations of the energy levels of minor trace elements are simulated using the Single Level Population Model (SLPM), which neglects interactions between different energy levels. This approach is suitable for resonant transitions that are primarily responsible for line cooling and heating through photoionization. To assess the role of heating or cooling due to the metals and to include this effect directly in the 3D simulations, we model the populations of the lowest energy levels of O I, C I, and Mg I, as well as the ions C II and Mg II, using the SLPM. The list of levels is specified in (Shaikhislamov et al. 2025). These elements are considered to be most important, especially for atmospheric cooling (Fossati et al. 2021; Huang et al. 2017; Nakayama et al. 2022). At the same time, Fe I and Fe II particles have too complex energy level structures for such an approach. The other notable cooling element, Na atom, is very rapidly photoionized and restricted to a narrow strip close to the planet (at pressures $> 10^{-6}$ bar, Huang et al. 2017) not affecting the absorption region.

To evaluate the validity of the SLPM approximation and to access the role of Fe I, Fe II in cooling of atmosphere, we employ a dedicated platform, ASTERA, which is analogous to CLOUDY and is being developed by us to simulate multilevel populations of elements (Shaikhislamov et al. 2025).

In what follows, the stellar spectral energy distribution (SED) is given at a reference distance of 1 au rather than at the planetary orbital distance. We divide the SED into HXR (0.5–10 nm), XRI (10–50 nm), EUV (500–91.2 nm), VUV (91.2–300 nm) and optical. The XUV flux of HAT-P-32 (0.5– 91.2 nm) is rather unconstrained because of lack of measurements in NUV range. Czesla et al. 2022 observed HAT-P-32 with XMM-Newton on 30 August 2019 and measured unusually high flux in the HXR range of 80 erg cm$^{-2}$ s$^{-1}$. The SED reconstructed by Czesla et al. 2022, using this measurement together with the model of Sanz-Forcada et al. 2011, contains the following fluxes (in erg cm$^{-2}$ s$^{-1}$): 80 in HXR, 110 in XRI, 315 in EUV, and $7\times10^4$ in VUV ranges. Thus, the total XUV range contains about 500 erg cm$^{-2}$ s$^{-1}$, which is about 100 times of solar value, an extremely large amount for an F-type star. Zhang et al. 2023 adopted a significantly more moderate SED based on calculations for similar star τ Boo, yielding fluxes of 32 in HXR and 89 in XRI, in erg cm$^{-2}$ s$^{-1}$.

We adopt even more moderate values for the stellar spectral energy distribution (SED) as our fiducial case: 15 in HXR, 67 in XRI, and 18 in EUV, resulting in a total XUV flux of 100 erg cm$^{-2}$ s$^{-1}$ at 1 au. Our fiducial flux in the Lyα line is 290 erg cm$^{-2}$ s$^{-1}$. The fiducial Lyα flux adopted in (Yan et al. 2024), based on measurement at ζ Dor, is 47 erg cm$^{-2}$ s$^{-1}$, which is about 6 times lower than the value adopted in the present work. However, the actual values used in the simulations of (Yan et al. 2024) are much larger, reaching at least ~500 erg cm$^{-2}$ s$^{-1}$. Thus, in the absence of suitable observational constraints, both the Lyα and XUV fluxes remain unconstrained parameters in the simulations. For this reason, we vary them in a wide range in order to check how they affect the calculated absorptions in the Hα and He I lines.

For the accurate comparison with the observations, we also checked the effect of phase smearing by averaging the synthetic absorption profiles over the corresponding part of the orbital phase (the 900 s exposure at the mid-transit used in the observations of Czesla et al. 2022). We find that this effect reduces the maximum absorption depth by less than 5%. Thus, we do not take it into account in this paper.

We note that the planetary equilibrium temperature is close to 2000 K, at which the dissociation of H$_2$ molecular atmosphere into atomic one takes place. It means that even small changes in energy balance can lead to widely different equilibrium states in the upper atmosphere. Because of this we simulate both cases of base atmosphere: a molecular one and an atomic one.

The Stellar Wind (SW) is modeled with the same H$^+$ and He$^+$ fluids as those describing the planetary components. It is launched through boundary conditions at the stellar corona and is driven by empirical heating term, as described in our previous papers (Khodachenko et al. 2019). The coronal conditions and the heating term are parameterized by values of the integral mass-loss rate and the asymptotic velocity of SW.

To investigate the absorption in spectral lines at mid-transit, in Sections 3.1 – 3.3 we simulate atmosphere within a sphere around the planet restricted by $R_{max} = 10R_p$, and we do not include the stellar wind. This simplification is sufficient for simulating the mid-transit absorptions, as the contribution from atmospheric material beyond this region is minimal when it is swept away by the SW. In Section 4, to model all transit features (pre, mid and post-transit), we solve equations in a global sphere including the star with the SW.

We begin presenting the results with a general picture of the simulations, shown as a 2D cuts of the 3D distributions of the most interesting species which produce observational signatures, as described in Sections 3 and 4. The spatial plots show the excited H I and He I and show how far they extend from the planet along with the planetary material. The flow streamlines and the structure of the upper atmosphere are very asymmetric and strongly skewed relative to planet-star line.

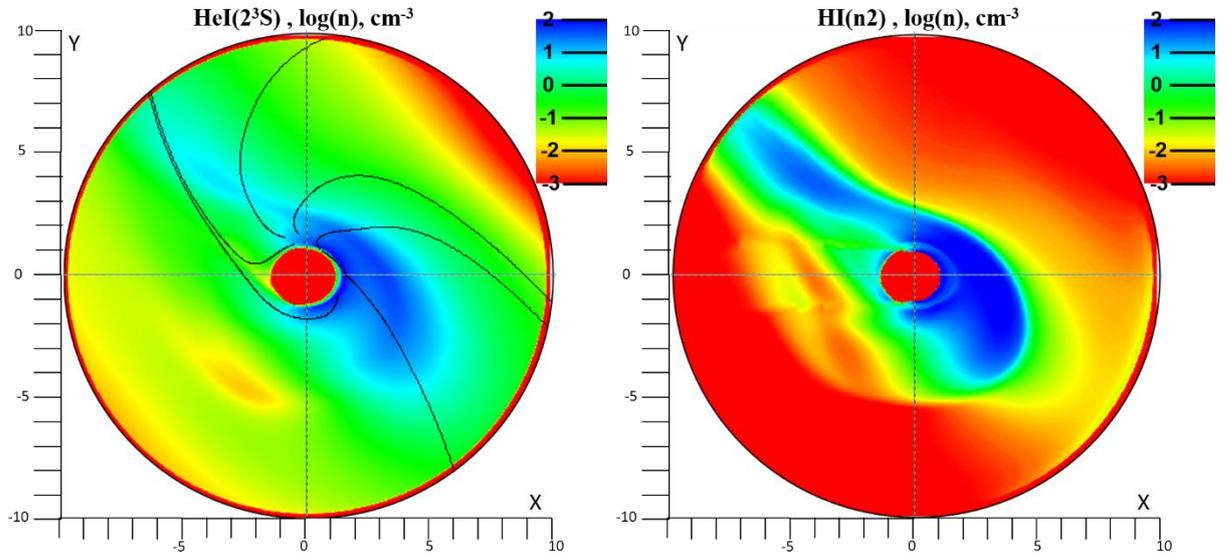

**Figure 1.** Color plots of the metastable helium density (left) and the excited hydrogen density (right) in the equatorial plane of the planet. The scale is in units of planetary radii. The star is located at $X = 40R_p$ on the right. The red circle indicates the planet. The black lines are streamlines of the atmospheric flow. Values outside the specified variation ranges are colored either in red if smaller than minimum, or in blue, if higher than maximum. Simulation parameters are given in Section 3.3 (run №23 in Table 3).

### 2.1 Scattering of Lyα photons

Diffusion of Lyα photons in both spatial and frequency domains is an important process, without which the population of excited H I cannot be calculated in atmospheric layers that are optically thick to Lyα. Such a treatment requires general kinetic modeling using Monte Carlo codes. However, in exponentially compact atmospheres, where the scale height is much smaller than the planetary radius, an effective Lyα photon loss rate can be employed, dependent on the local value of the absorption parameter $\alpha = n_{HI}\sigma H$, where $n_{HI}$ is the density of H I in the ground state, $H(T)$ is the local scale height, and $\sigma$ is the absorption cross-section averaged over the thermal Doppler width at the Lyα line center. This approach allows to calculate the population of excited hydrogen within a local approximation. We have tested this method for the atmosphere of KELT-9 b (Shaikhislamov et al. 2025). For HAT-P-32 b, we apply it similarly, comparing the results from hydrodynamic and Monte Carlo simulations. The Monte Carlo code developed by our team is described in (Miroshnichenko et al. 2021; Sharipov et al. 2023). For direct comparison, we simulated atmosphere without tidal forces, so as to make it rotationally symmetric around planet-star line (the geometry of the employed Monte Carlo code), and without stellar Lyα flux, so that only the diffusion of locally generated photons is considered. The resulting spatial distributions of hydrogen atoms, protons, electrons, and temperature were used as the atmospheric structure for the Monte Carlo simulations.

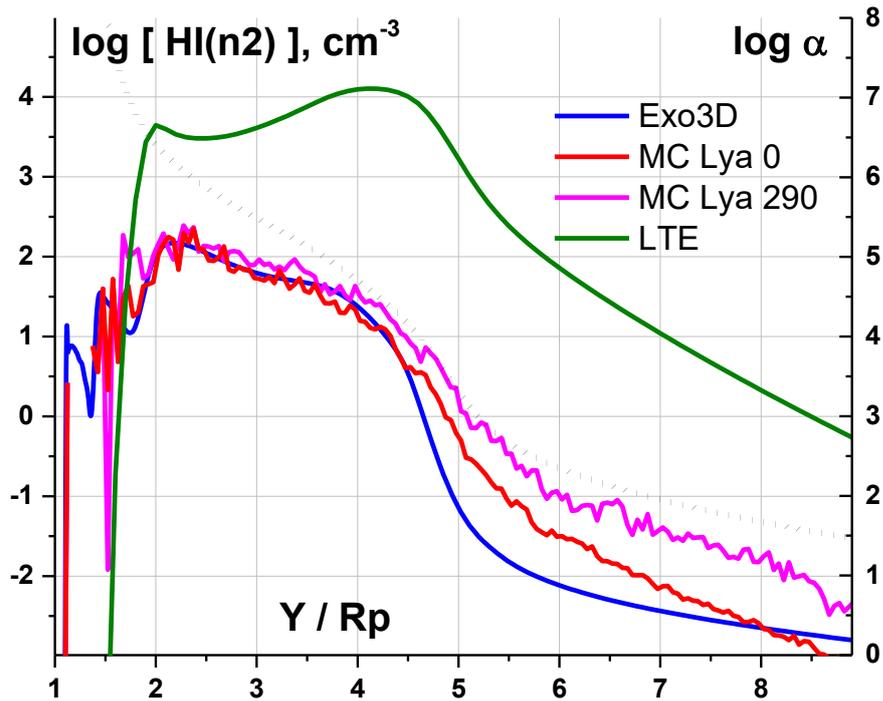

**Figure 2.** Profiles of hydrogen excited to the $n = 2$ level, calculated by hydrodynamic code (Exo3D, blue line) and by Monte Carlo code (MC, red). Parameters of simulations are the same, with zero stellar Lyα flux. Also is shown the MC calculation with fiducial Lyα stellar flux (magenta). Olive line shows LTE population of H I($n$=2) level. The right axis and black dotted line indicate the absorption parameter, log(α).

The result of such modeling is shown in the Fig. 2. One can see that agreement between full MC and local HD simulations is very good in regions where the Lyα photons are strongly trapped (absorption parameter α > 1000, r < 5R$_p$). As will be shown later, the Hα absorption comes namely from this region, while the relatively rarified atmosphere above 5R$_p$ does not produce any effect. Nevertheless, Lyα photons are still absorbed at R > 5R$_p$; they cannot penetrate to deeper layers, and this self-shielding effect explains why the stellar Lyα flux has a relatively small effect on Hα absorption, as will be demonstrated later. Fig. 2 verifies this through MC simulation with the fiducial Lyα flux. There is also shown the LTE population of H I(*n*=2) level which verifies that the non-LTE kinetic simulation is indeed necessary.

### 3.1 Atomic atmosphere

Table 1 lists the simulation runs with different input parameters and main output characteristics. Figures 3 and 4 show mid-transit absorption profiles at different values of helium abundance, base temperatures and SED characteristics across different spectral ranges. One can see that the He I line is sensitive to VUV flux. It means that photoionization is an important process for depopulating the metastable level. On the other hand, the Hα line is not affected by the VUV flux but is sensitive to HXR flux, similar to the He I line. This is because HXR penetrates deep into the dense atmosphere, effectively heating and expanding it. There is a moderate influence of stellar Lyα flux on the Hα line, and a negligible effect on the He I line. In the absence of Lyα flux, the Hα absorption is reduced by a factor of two.

**Table 1.** List of simulation runs for the atomic atmosphere. From left to right: assumed parameters: helium to hydrogen abundance, integral fluxes in the XHR, VUV parts of the SED, and Lyα line in erg cm$^{-2}$ s$^{-1}$ at 1 au; derived values: atmospheric total mass-loss rate in $10^{10}$ g s$^{-1}$, amplitude of absorption in mid-transit and equivalent width in Å in Hα and He I lines. Last column shows other parameters. The first row gives values measured in observations. The notation "×M" means that the fiducial value is multiplied by a factor of M. Blanc cells have the same value as the last above in the column, except "×M" notations which are not repeated.

| № | H/He | HXR | VUV | Lyα | M$^/$ | Hα | EW | He I | EW | comment |
|---|------|-----|-----|-----|-------|-----|-----|------|-----|---------|
| 1 |      | 80  |     |     |       | 0.057 | 0.046 | 0.064 | 0.093 |         |
| 2 | 91/9 | 14.5 | 6.6·10$^5$ | 290 | 815 | 0.033 | 0.022 | 0.093 | 0.154 | T$_{base}$ = 0.18 |
| 3 | 93/7 |     |     |     | 934   | 0.036 | 0.026 | 0.083 | 0.149 |         |
| 4 | 96/4 |     |     |     | 1300  | 0.04  | 0.042 | 0.058 | 0.132 |         |
| 5 | 91/9 |     |     |     | 650   | 0.028 | 0.016 | 0.088 | 0.124 | T$_{base}$ = 0.15 |
| 6 | 98/2 |     |     |     | 937   | 0.049 | 0.037 | 0.033 | 0.128 | T$_{base}$ = 0.20 |
| 7 | 91/9 | ×2  |     |     | 1130  | 0.045 | 0.036 | 0.114 | 0.226 | T$_{base}$ = 0.18 |
| 8 |      |     | ×2  |     | 820   | 0.033 | 0.021 | 0.068 | 0.101 |         |
| 9 |      |     |     | ×0  | 744   | 0.02  | 0.013 | 0.09  | 0.14  |         |
| 10 |     |     |     | ×2  | 917   | 0.039 | 0.028 | 0.095 | 0.175 |         |
| 11 |     | ×2  | ×2  |     | 1550  | 0.052 | 0.054 | 0.087 | 0.198 | T$_{base}$ = 0.20 |

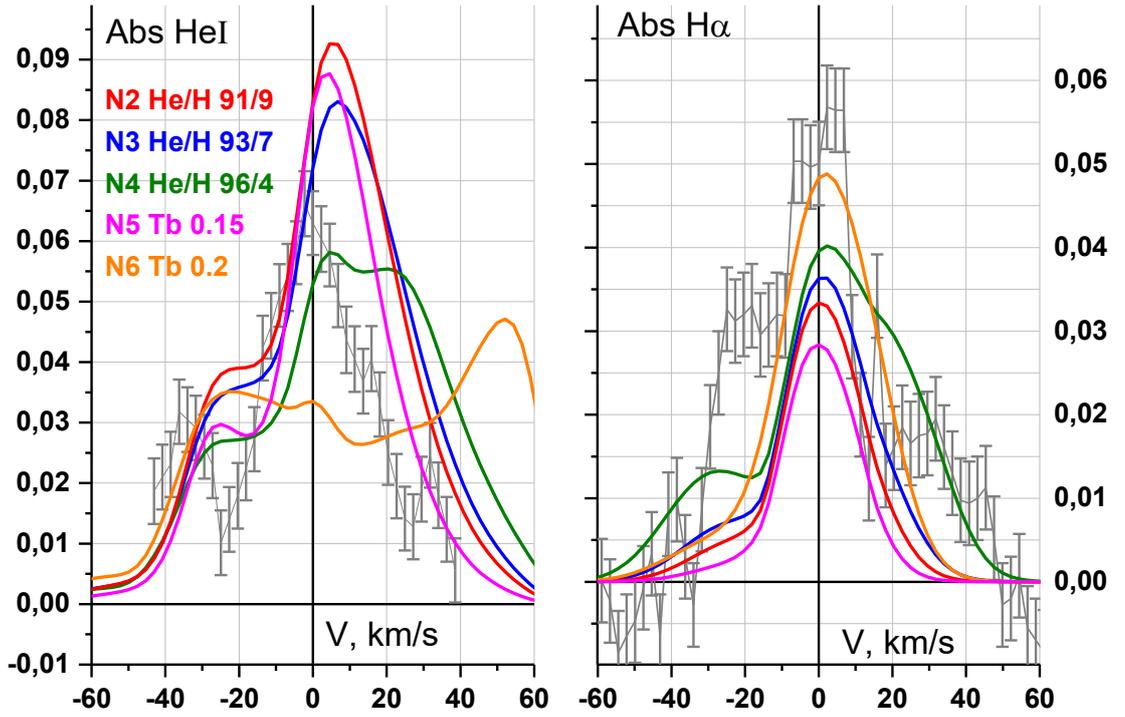

**Figure 3.** Simulated absorption profiles of the He I 1083 nm line (left panel) and the Hα line (right panel) in Doppler velocity units. The model runs corresponding to the Table 1 list and some of the parameters are given in the legend by different colors. Here and further gray points with error bars show the measurements from (Czesla et al. 2022).

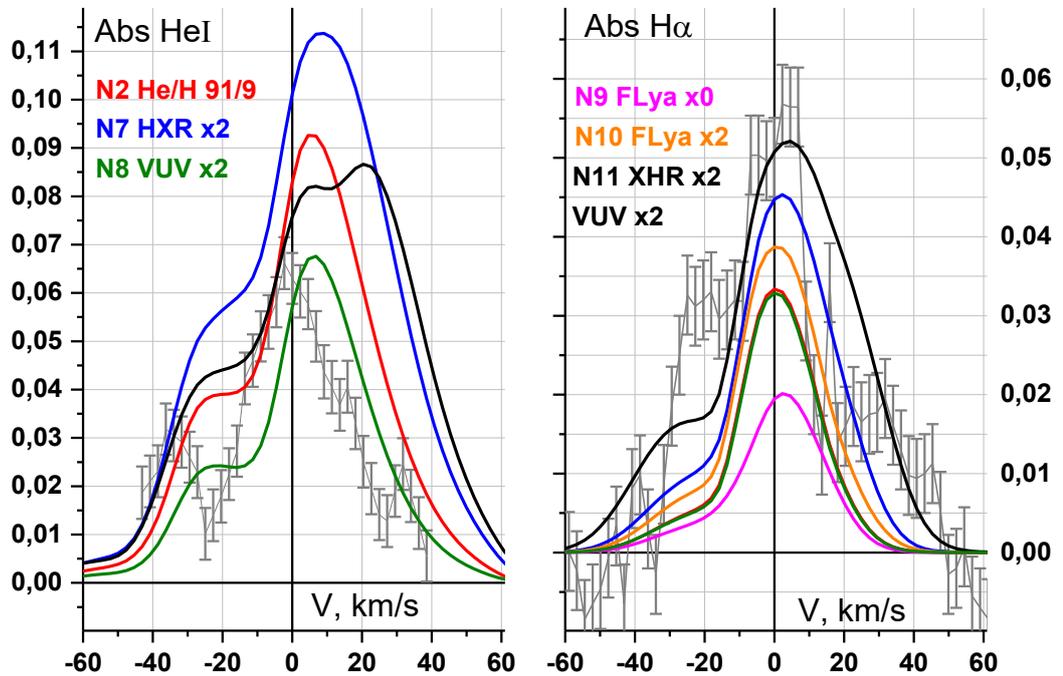

**Figure 4.** Simulated absorption profiles at different SED parameters. In the He I plot, curves №9, №10 are not shown because they are very close to the №2 curve.

It appears that the pure atomic atmosphere of HAT-P-32 b is extremely inflated and remains dense even beyond the Roche lobe. Because of that, even at relatively low fiducial HXR flux and an XUV flux of 100 erg cm$^{-2}$ s$^{-1}$, the absorption is sufficiently large to match the measured depth in both lines. However, due to the relatively high velocities, this dense plasma produces strong Doppler broadening in both absorption lines. Such broadening does not fit the mid-transit observations at all at moderate helium abundances H/He > 95/5, especially in the red wing of the He I line, which is produced by the part of the flow directed toward the star. At H/He ≈ 92/8, the excess absorption in He I is too large, while in Hα line too small.

Adding metals drastically increases the outflow rate. The reason is that photoionization of elements with low ionization potential (Mg, Fe) supplies electrons in the dense layers of atmosphere, increasing overall heating. Due to expansion of atmosphere the absorption in lines takes place further away from the planet, resulting in pronounced blue and red-shifted wings. This is demonstrated in Fig. 5. One can see that even at a metallicity of only tenth of the solar value, the absorption profiles still do not fit observations in the He I line. Thus, our modeling predicts that a purely atomic He/H atmosphere cannot fit the observations, regardless of metallicity, due to the large Doppler broadening, especially in the He I line.

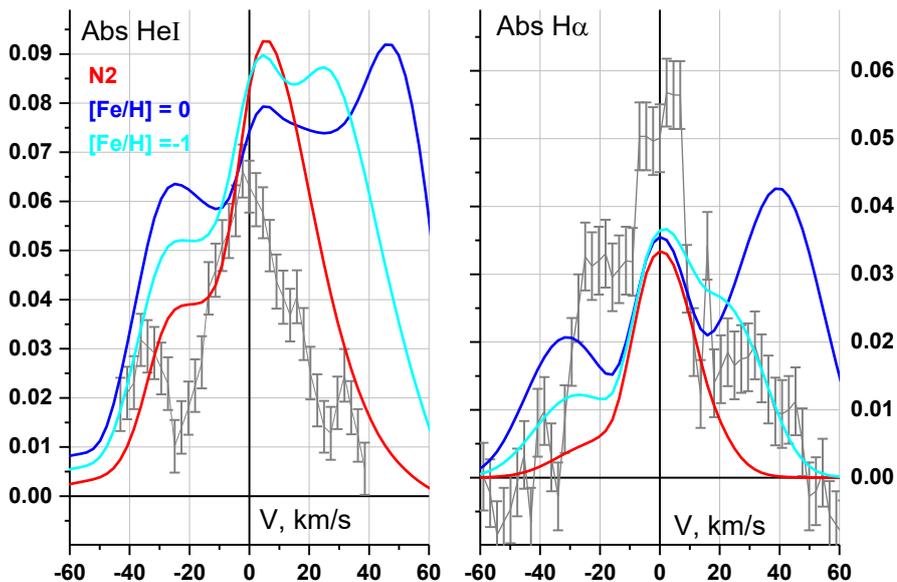

**Figure 5.** Simulated absorption profiles with solar value (blue) and 1/10 of solar value (cyan) metallicity. Red line shows the case with absence of metals.

### 3.2 Molecular atmosphere

First, we consider a pure molecular $H_2$ + He atmosphere without metals. Table 2 lists the parameters of the corresponding simulation runs. Due to the approximately two times smaller scale height, such an atmosphere is significantly more compact. For the fiducial SED, the modeled line absorption is significantly lower than the observed values, especially for Hα. Assuming a high level of

$F_{HXR}$ = 80 erg cm$^{-2}$ s$^{-1}$, as measured by XMM-Newton, the absorption significantly increases. However, for Hα it still remains below the observed value even for an increased base temperature of 2000 K. Additional increasing of high-energy photon stellar flux in the XRI range by 50% affects absorption very slightly. The absorption in the He I line can be tuned by varying the VUV flux, as demonstrated in Fig. 7.

**Table 2.** List of simulation runs for a molecular atmosphere. Notations are the same as in Table 1.

| № | H/He | HXR | VUV | Lyα | M$^/$ | Hα | EW | He I | EW | Comment |
|---|---|---|---|---|---|---|---|---|---|---|
| 1 | | **80** | | | | 0.057 | 0.046 | 0.064 | 0.093 | |
| 12 | 95/5 | **14.5** | 6.6·10$^5$ | 290 | 230 | 0.005 | 0.002 | 0.024 | 0.025 | **T$_{base}$ = 0.18** |
| 13 | | **80** | | | 450 | 0.19 | 0.012 | 0.05 | 0.056 | |
| 14 | | **14.5** | | | 365 | 0.13 | 0.007 | 0.039 | 0.043 | **T$_{base}$ = 0.20** |
| 15 | | **80** | | | 860 | 0.045 | 0.033 | 0.081 | 0.117 | |
| 16 | **98/2** | | | | 937 | 0.049 | 0.037 | 0.052 | 0.07 | |
| 17 | | | | | 960 | 0.052 | 0.039 | 0.055 | 0.069 | **XRY ×1.5** |
| 18 | | | **×0.5** | | 937 | 0.05 | 0.037 | 0.071 | 0.104 | |
| 19 | | | | **×3** | 1000 | 0.057 | 0.044 | 0.054 | 0.074 | |
| 20 | | | | **×6** | 1120 | 0.062 | 0.051 | 0.055 | 0.077 | |

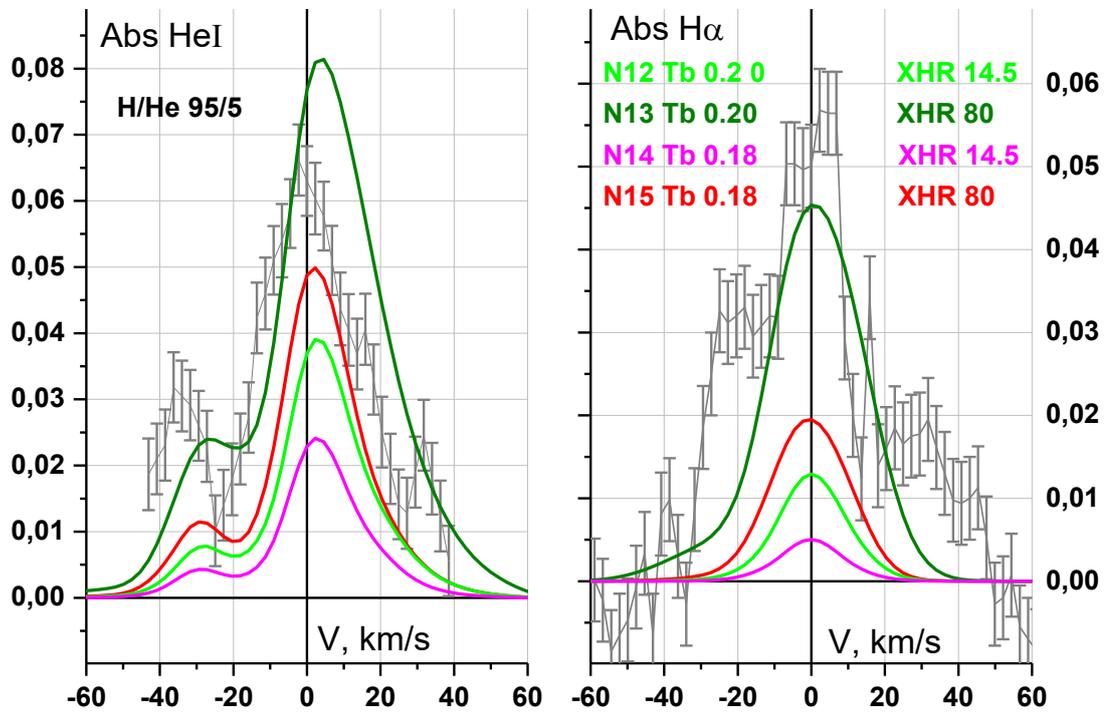

**Figure 6.** Simulated absorption profiles for different HXR fluxes and base temperatures. The helium abundance is the same for all runs, with He/H = 0.05.

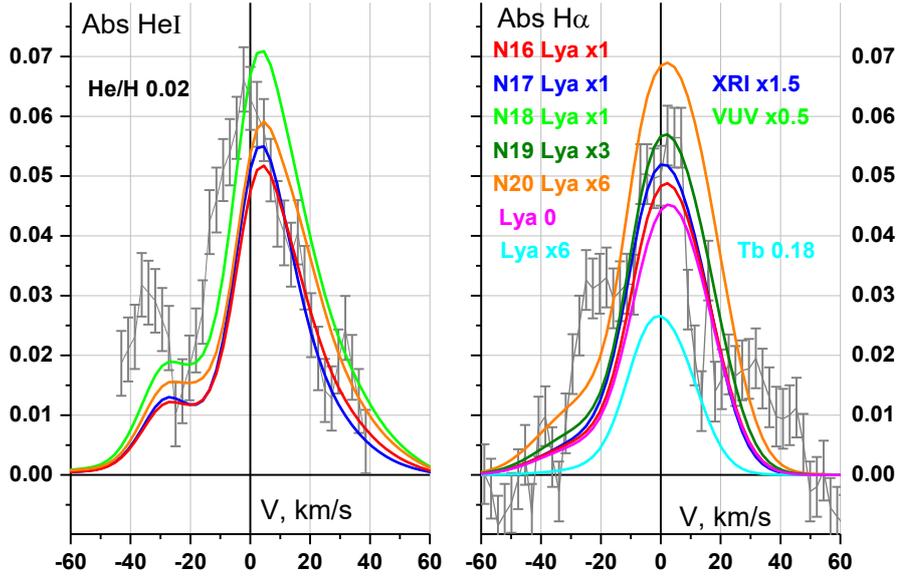

**Figure 7.** Simulated absorption profiles for different SED parameters and Lyα fluxes. The following simulations are shown separately: the case with zero Lyα flux (magenta); the case with a base temperature of 1800 K and a sixfold increase of Lyα flux (cyan), while other parameters are the same as in run №16.

The fit to the Hα is obtained with a significant severalfold increase in the Lyα flux. From simulations performed under the conditions of run №16, it was found that the dependence of the maximum Hα absorption on the Lyα flux can be parametrized as $A_{H\alpha} = 0.0448 + 0.004 \times (F_{Ly\alpha}/F_{Fid})$, where the fiducial flux is 290 erg cm$^{-2}$ s$^{-1}$. One can see that the effect of stellar Lyα photons on the excitation of H I($n=2$) is very moderate. This is due to the efficient shielding of Lyα photons in the relatively rarified atmosphere, which does not make the largest input into the absorption, as demonstrated in Fig. 1. To fit the observations, a very large Lyα flux of about 800 erg cm$^{-2}$ s$^{-1}$ is required. This is similar to the conclusion derived in (Yan et al. 2024). We note that for both lines the best fit is obtained at $T_{base} = 2000$ K. At a base temperature of $T_{base} = 1800$ K, no fit to the Hα line can be achieved.

### 3.3 Molecular atmosphere with metals

The presence of metals significantly affects the upper atmosphere and absorption by supplying electrons in deeper layers, where radiation capable of ionizing hydrogen does not penetrate. Free electrons facilitate more efficient dissociation of H$_2$ which leads to a more extended atmosphere. The effect of metallicity on the line absorption is directly demonstrated in Fig. 8. The dependence of the absorption maximum and full width is shown in Fig. 9. One can see that even a metallicity of 0.01 Solar already produces increase in absorption, while at [Fe/H] = −1 the absorption is several times larger than in the metal-free case. The most rapid increase occurs in the range [Fe/H] = [−2, −1] for both lines. Fig. 10 shows the profiles of the main atmospheric components for Solar and very low metallicities, revealing an increase in H I in the region where the Hα and He I absorption originates, as well as an overall increase in electron density, especially close to the planet. The densities of excited H I and He I both show a strong increase at $R > 2R_p$ when metals are present in the atmosphere.

**Table 3.** List of simulation runs for a molecular atmosphere with metals. The notations are the same as in Table 1. The second column shows the logarithm of the metallicity relative to Solar. (*) The last two rows correspond to cases in which the excited levels of C, O, and Mg are not calculated. The simulation without excited levels of H I is made only in run №25.

| № | Fe/H | He/H | HXR | Lyα | M′ | Hα | EW | He I | EW | comment |
|---|---|---|---|---|---|---|---|---|---|---|
| 1 | | | **80** | | | **0.057** | **0.046** | **0.064** | **0.093** | |
| 21 | −3 | 0.02 | 80 | 290 | 370 | 0.013 | 0.007 | 0.023 | 0.021 | $T_{base} = 0.18$ |
| 22 | −1.5 | | | | 550 | 0.025 | 0.015 | 0.035 | 0.036 | |
| 23 | 0 | | | | 935 | 0.043 | 0.03 | 0.052 | 0.065 | |
| 24 | | | | | 1360 | 0.06 | 0.052 | 0.062 | 0.094 | $T_{base} = 0.20$ |
| 25 | | | | | 1010 | − | − | 0.037 | 0.050 | H I($n = 1$) |
| 26 | | | | ×0.25 | 1370 | 0.57 | 0.049 | 0.062 | 0.094 | |
| 27 | | | | ×0 | 1100 | 0.051 | 0.044 | 0.062 | 0.094 | |
| 28 | | | | **290** | 1390 | 0.073 | 0.062 | 0.064 | 0.095 | * |
| 29 | | | | ×0 | 1380 | 0.069 | 0.059 | 0.063 | 0.094 | * |

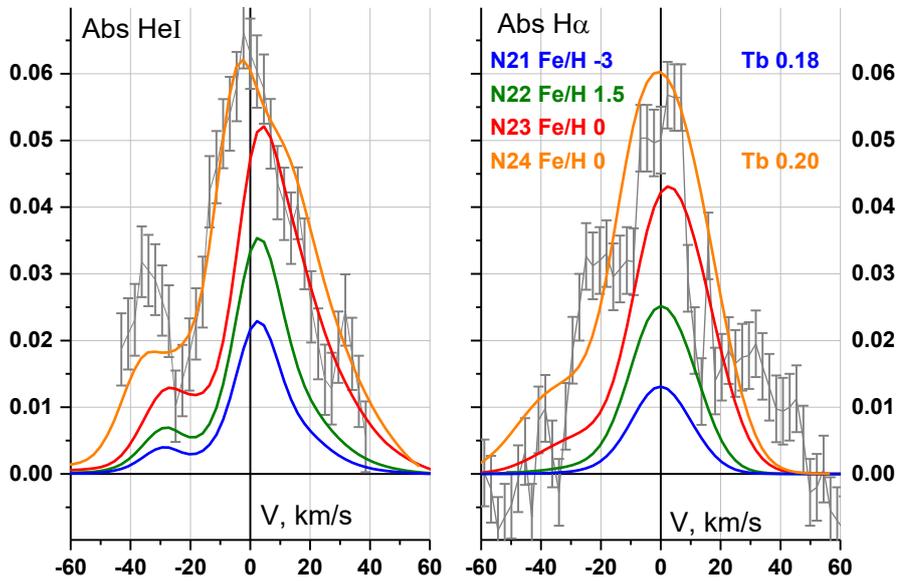

**Figure 8.** Simulated absorption profiles for different metallicities calculated for an atmosphere with a base temperature of 1800 K. Also is shown the calculation with $T_{base} = 2000$ K at Solar metallicity. For this particular case the Hα and He I profiles are shifted by $-7$ km s$^{-1}$, which provides the best fit.

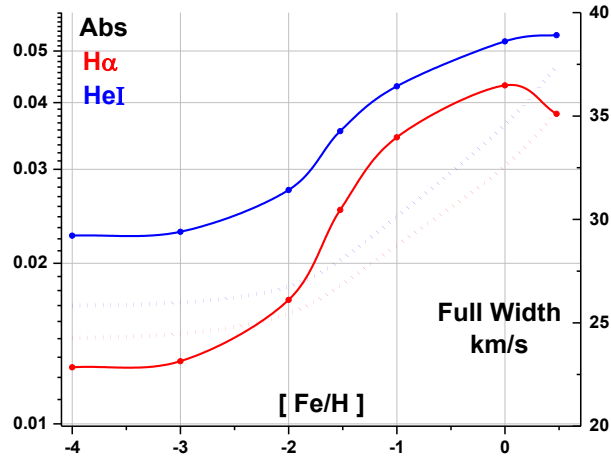

**Figure 9.** Dependence of the absorption maximum (solid lines, left axis) and the full width (dotted, right axis) on metallicity for the Hα (red) and He I (blue) lines. The other parameters are the same as in the run №23.

To demonstrate the distribution of absorption in the atmosphere, we plot it in Fig. 11 as a function of the cut-off radius. That is, absorption is not included if a point along the LOS lies outside the specified region, $R > R_{cut}$. One can see that approximately half of the absorption for both lines originates from layers below $2.5R_p$, while about 90% of the absorption is restricted by heights $(3.5–4.5)R_p$. The right panel of Fig. 11 shows the integrated evaluation of the local optical depth in dependence on radius. It shows that, for the considered lines, the atmosphere is optically thin above $\sim 2R_p$, while below this level it becomes moderately optically thick.

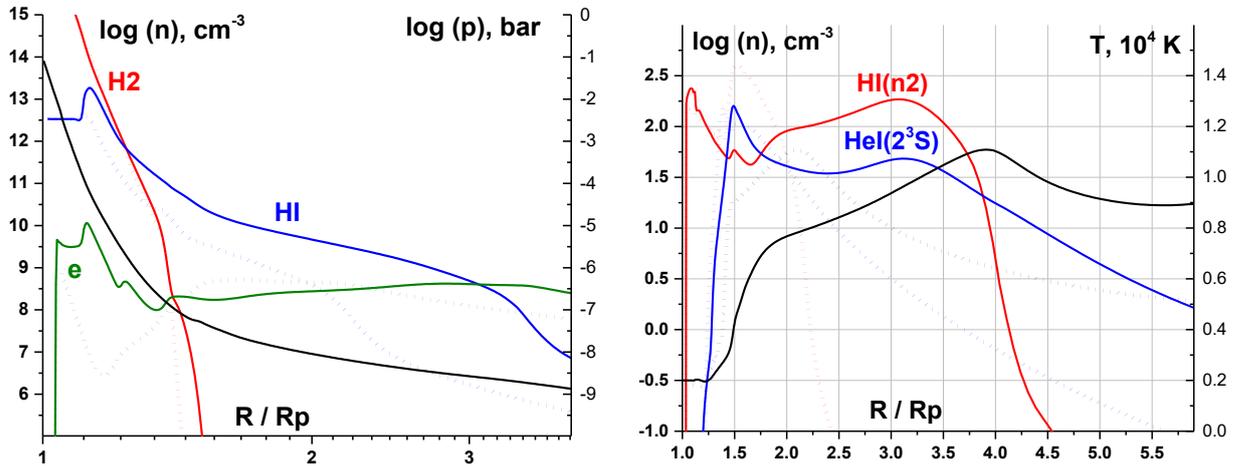

**Figure 10.** Profiles of main values along the planet-star line for cases of Solar metallicity [Fe/H] = 0 (solid lines), and very low metallicity [Fe/H] = −4 (dotted lines). The other parameters are the same as in run №23.

Left panel: densities of molecular hydrogen (red), atomic hydrogen (blue), and electrons (olive). The right axis shows the pressure profile (black line).

Right panel: densities of excited hydrogen (red) and metastable helium (blue). The right axis shows the temperature profile (black lines).

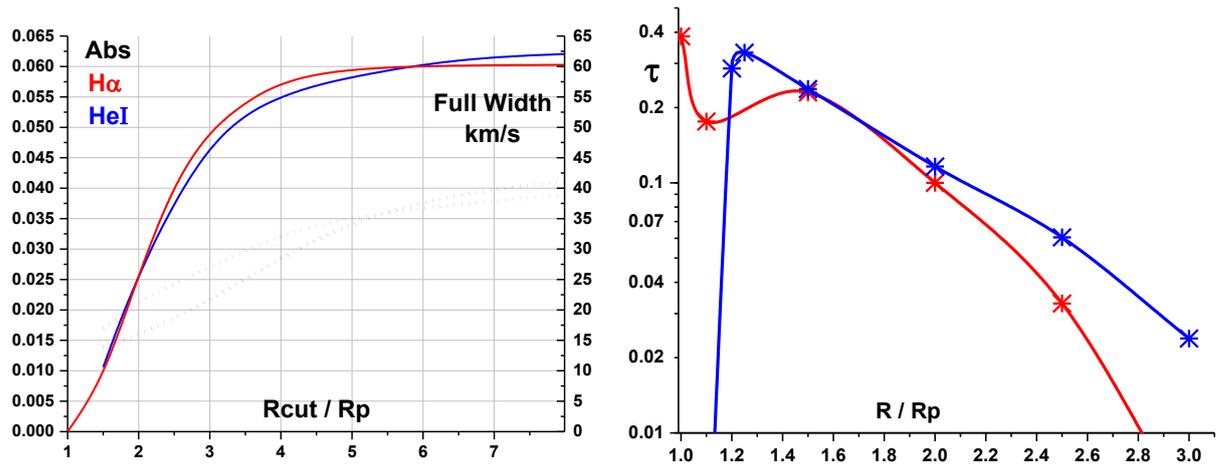

**Figure 11.** Left panel: Absorption maximum (left axis, solid lines) and full width (right axis, dotted lines) calculated within the volume of atmosphere confined by the maximum radius $R_{cut}$. Hα – red, and He I – blue lines. Simulation run №23.

Right panel: Optical depth calculated for the specified spherical shell $\Delta = 0.1R$ as a function of radius R.

For the parameters that provide a relatively good fit to the observations, we adopt HXR = 80 erg cm$^{-2}$ s$^{-1}$, as measured by XMM-Newton, a relatively low helium abundance (He/H = 0.02), and Solar metallicity. The Hα absorption is weakly sensitive to the Lyα flux, which can be in the range of 50–500 erg cm$^{-2}$ s$^{-1}$. For the profiles shown in Fig. 10, the optical depth at the center of the Lyα line is about $10^4$ at a distance of $R = 4R_p$. That is, the region where most of the Hα absorption occurs (see Fig. 11) is fully shielded from Lyα photons. An increase in the Lyα flux affects only region relatively far from the planet, adding little to the overall absorption.

The derived mass-loss rate for these parameters is about $10^{13}$ g s$^{-1}$. Interestingly, this value is derived by several rather different models cited in Introduction, which were constrained by fitting the absorption depth of the He I line, or both lines. The present results show a persistent shift of about 7 km s$^{-1}$ of the synthetic profiles relative to the observations. It is generated by a portion of the atmosphere moving toward the star beyond the Roche lobe and twisted by the Coriolis force. In view of the significant pre- and post- transit absorptions measured in both lines, this discrepancy may indicate a complex picture of the observed absorptions, consisting of at least two different structures: one in the relative vicinity of the planet, and another extending far ahead of and behind the planet, or even encompassing the star.

Considering the best-fit solution (run №24), we demonstrate in Fig. 12 the importance accounting for the kinetics of excited levels. Run №25 was performed without excited levels of H I. In this case, the Hα absorption can be calculated assuming LTE, yielding an amplitude of about 0.5, which is unrealistically larger than the measured value. But the key point here is that excited levels of hydrogen increase the effective heating of the atmosphere; without this effect, the helium absorption is almost twice smaller than observed. Without the kinetics of metals

(run №28), the cooling due to excitation of their low levels is neglected, resulting in higher atmospheric temperatures and, consequently, stronger Hα absorption. The relative unimportance of the stellar Lyα flux in populating the H I($n$=2) level is demonstrated in the run №26. Reducing it by a factor of 4, to 73 erg cm$^{-2}$ s$^{-1}$, does not affect the He I line, while the Hα absorption decreases by mere 5%.

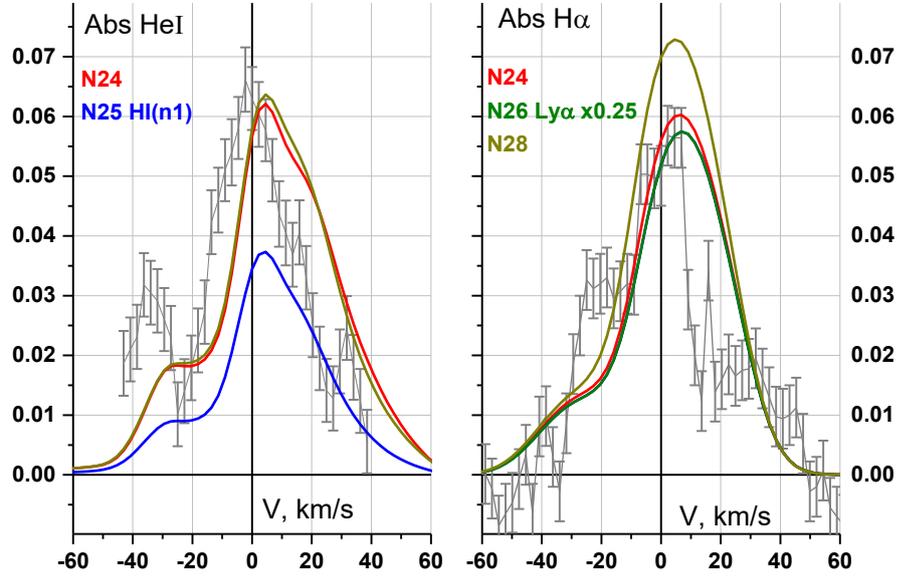

**Figure 12.** Simulated absorption profiles calculated for an atmosphere with a base temperature of 2000 K and Solar metallicity (run №24). For comparison, there are shown the case of 4-times lower stellar Lyα flux (run №26), the case without calculation of excited hydrogen populations (run №25), and the case without calculation of excited levels of metals (run №28).

### 3.3 Formation of leading and trailing tails

For most hot exoplanets, a planetary atmosphere outflowing beyond the Roche lobe at a rate of $\leq 10^{11}$ g s$^{-1}$ is swept away from the star even by a weak plasma wind. However, the huge mass loss of HAT-P-32 b is actually comparable to, and even higher than the Solar average mass-loss rate. Because the thermal and kinetic energies of escaping particles, < 10 eV, are much smaller than the gravitational well of the star, the planetary material cannot escape the orbit on its own and instead spreads along the orbit. Without sufficient pressure from the SW, the leading stream gradually falls toward the star, while the trailing tail gradually moves outward. Observations by (Zhang et al. 2023) support this scenario, revealing in the He I line absorption beyond the optical egress and unusually deep and extended ingress. The projection of the leading tail covers about 2 stellar diameters. 3D HD simulations by Zhang et al. 2023 demonstrated formation of a torus of planetary material encircling the star, although these simulations did not fully reproduce the duration of the ingress. 3D HD simulations by (Nail et al. 2025), aimed to achieve the asymmetric light curve with extended ingress, required rather cool atmospheric temperature (T ≈ 4700 K) with the nightside flow suppressed relative to the dayside.

We performed simulations in a global reference frame including the star and the SW, using the parameters identified above as providing the best fit to the mid-transit absorption. For parameter set №24, the Solar SW (integral mass-loss rate $3\times10^{12}$ g s$^{-1}$ = $4.7\times10^{-14}$ M$_{sun}$) carries away the leading tail, whereas a SW about order of magnitude weaker is pushed out by planetary material from the equatorial plane.

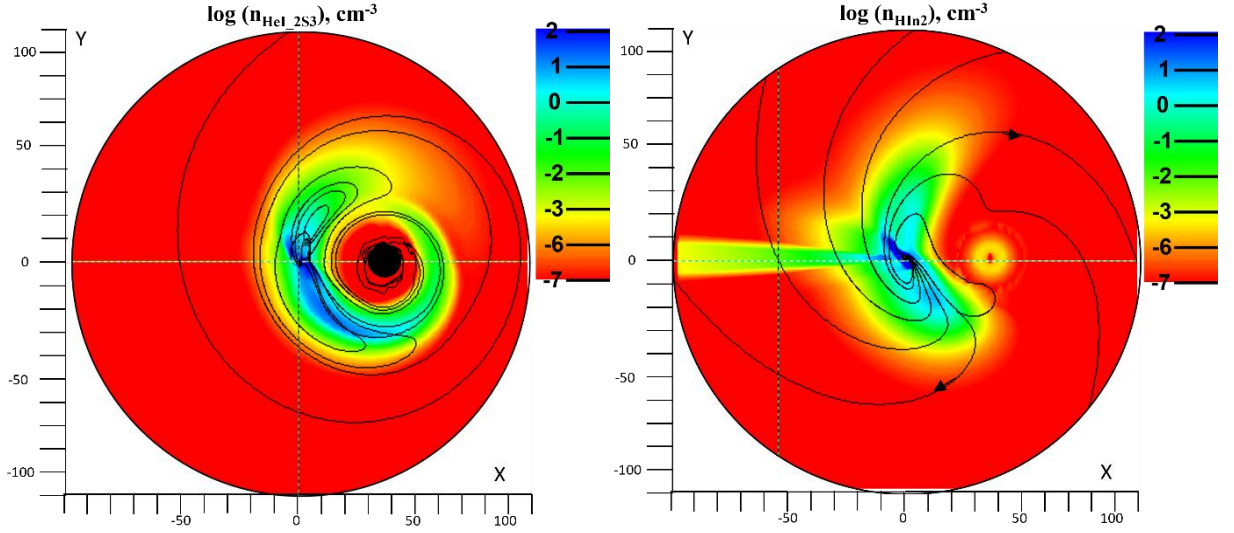

**Figure 13.** Color plot of metastable helium and excited hydrogen atoms in the equatorial plane of the planet, simulated with the parameter set №24 (Table 3). Left panel: weak SW ($M'_{sw} = 10^{11}$ g s$^{-1}$, density of metastable helium He I($2^3$S)). Right panel: moderate SW ($M'_{sw} = 3\times10^{12}$ g s$^{-1}$, density of excited hydrogen H I($n=2$)). Black lines indicate streamlines of atmospheric flow. The black circle shows the star to scale.

In case of a weak SW, planetary material forms a belt around the star, with the leading tail dominating and gradually falling toward the star. This results in a strongly asymmetric transit curve, with the ingress much more extended than the egress, as shown in Fig. 14. Note that the case of a moderate SW, as well as the simulation performed in a region of 10R$_p$ around the planet without including the star and SW (run №24), produce nearly symmetric transit curves.

The material accumulating around the star affects the mid-transit absorption, as shown in Fig. 15. This material does not share the planetary RV (Radial Velocity), whereas the material close to the planet does. These different streams separated in space produce significantly broader absorption at mid-transit than is observed. We note that the transit curves in Fig. 14 are calculated in the stellar frame, consistent with the observational data of (Zhang et al. 2023). The absorption profiles in Fig. 15 are calculated in the planetary frame, also consistent with observations, in this case from (Czesla et al. 2022). Although we do not accurately fit at the same time the full transit curve and the mid-transit line profile, it can be seen that the best fit for both datasets lies between the moderate and weak SW cases.

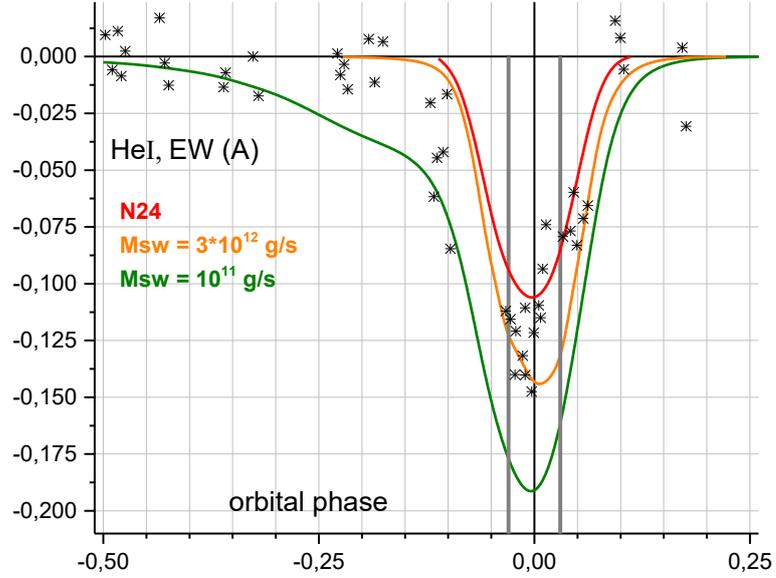

**Figure 14.** Absorption transit curves in the He I line, averaged over the interval ±50 km s$^{-1}$ and expressed as equivalent width, for the cases of weak (olive line) and moderate (orange) SW. The other parameters are the same as in run №24. The simulations are performed in a global frame including the star, whereas run №24 was performed in a sphere including only the planet. Gray vertical lines denote optical transit window. Crosses show the observational data from (Zhang et al. 2023).

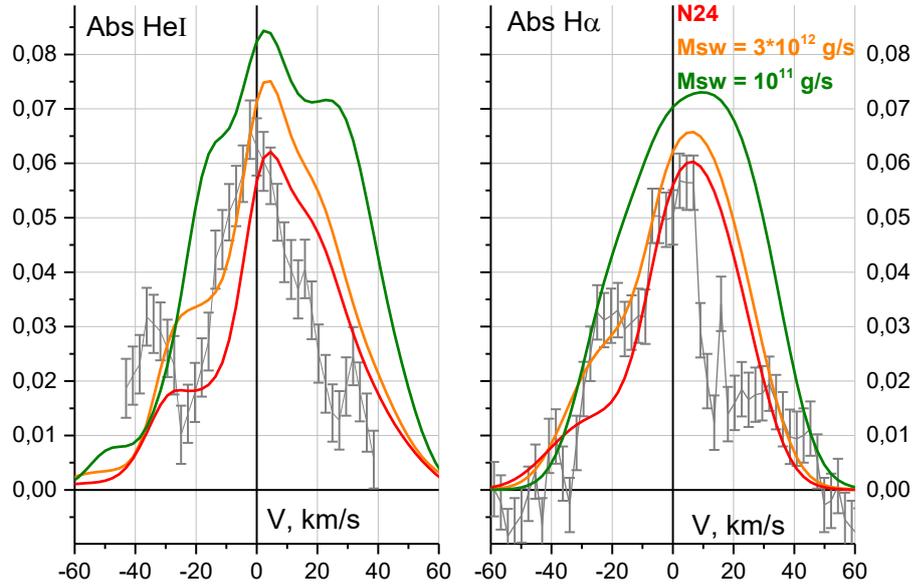

**Figure 15.** Absorption profiles at mid-transit calculated for weak (olive curves) and moderate (orange) SW.

## Conclusions

We have modeled the atmosphere of HAT-P-32 b, including absorption in the Hα and He I lines, increasing the complexity of the simulations. First of all, it appears that achieving the absorption depth of 5.5% in Hα line at mid-transit requires a base temperature of 2000 K at pressures of ~0.1 bar, which is somewhat larger than the estimated equilibrium temperature of 1800 K. Especially it is apparent for the atmosphere of molecular $H_2$. Even this relatively small increase in base temperature raises the overall atmospheric density, resulting in a strong enhancement of the Hα absorption. We note that other simulations of HAT-P-32 b cited above start their atmospheric models from much lower pressures. Despite this, for a characteristic density of H I atoms $3×10^{11}$ cm$^{-3}$, the corresponding heights and temperatures are: R ≈ 1.25×$R_p$, T ≈ 6×$10^3$ K in simulations of (Czesla et al. 2022); R ≈ 1.1$R_p$, T ≈ 3×$10^3$ K in Yan et al. 2024, and R ≈ 1.3$R_p$, T ≈ 3×$10^3$ K in the present work. Thus, regarding the regions where the absorption occurs, the typical conditions of different models are not significantly apart from each other.

The simplest case of an atomic atmosphere exhibits the largest mass-loss rate, up to $1.5×10^{13}$ g s$^{-1}$. Absorption depth in the Hα and He I lines are comparable to or higher than the measured values are achieved at HXR fluxes much lower than those measured by XMM-Newton. However, the absorption profiles, especially of He I, show too large broadening in the red wing of the lines, reflecting the outflow of dense stream ahead of the planet. Including metals increases this discrepancy with the observations, even at a metallicity of 10% Solar.

A pure molecular atmosphere produces absorption that is too low. To match observations, the helium abundance must be low (H/He ≈ 98/2), the HXR flux must be as high as measured by XMM-Newton. Besides, to reproduce the Hα line depth, the Lyα flux must be very high (~1000 erg cm$^{-2}$ s$^{-1}$ at 1 au). These results are in good agreement with the conclusions of (Yan et al. 2024).

Finally, adding metals to a molecular atmosphere accelerates dissociation of $H_2$, producing results that lie intermediate between pure H and pure $H_2$ cases. Metals begin to influence line absorption at abundance of about $10^{-3}$ Solar, reaching a maximum at about Solar metallicity. The physical mechanisms underlying this influence are twofold: metals supply electrons deep in the atmosphere, while also contributing to cooling. The cooling reduces the Hα absorption by roughly 15% and must be taken into account, as was demonstrated in a number of works where kinetics of excited levels of the most important elements was calculated (Fossati et al. 2021; Huang et al. 2017). Another important point is that heating associated with excited levels of H I is rather important for the absorption in the He I line.

One of the main findings is that, for Solar metallicity, it is not necessary to assume an extreme Lyα flux to fit the observed Hα absorption. Moreover, the dependence on the Lyα flux is relatively week due to self-shielding by atmospheric layers located farther from the regions where the Hα absorption actually occurs.

We also simulated the problem in a global reference frame including the star and its plasma wind. As found in previous simulations of (Zhang et al. 2023) and (Nail et al. 2025), the planetary outflow is strong enough to form a large orbital structure around the star. In the simulations cited above, a SW of about Solar intensity cannot sweep away this structure. In the present work, a SW with the Solar integral mass-loss rate sweeps away the planetary material streaming ahead of the planet. In this case, the line absorption is similar to that obtained in calculations without a SW. Conversely, a SW an order of magnitude weaker cannot sweep away the planetary tails, which form a spiraling torus around the star. This torus is asymmetric, producing asymmetric transit curve with a much more pronounced pre-transit absorption than post-transit absorption, consistent with observations. We note that in our simulations there is no need to assume some specific features of the planetary outflow, like domination of the day-side outflow and rather cold thermosphere (Nail et al. 2025). Also, we fully model the generation of planetary wind with realistic adiabatic specific heat ratio of $\gamma = 5/3$, rather than launch it at an assumed temperature and mass-loss rate and calculating its dynamic with $\gamma \approx 1$. We do not simultaneously fit the mid-transit line profiles and the transit curves exactly. However, the important constraints and physical features derived in observations, such as the maximum depth and equivalent width of both lines, the asymmetric transit curve in the He I line, can be matched by our simulations, allowing robust conclusions to be drawn. These conclusions are: (1) the hard X-ray flux of the star at the time of observations of the H$\alpha$ and He I lines should be similar to that measured by XMM-Newton several months later; (2) the metallicity of atmosphere should be about the Solar; (3) there is no need to invoke an extreme stellar Ly$\alpha$ flux to fit the depth in H$\alpha$ line; (4) the plasma wind of the star should be most likely weaker than the Solar Wind.